\documentclass[prb,twocolumn,superscriptaddress,nopacs,preprintnumbers,amsmath,amssymb]{revtex4}
\usepackage{graphicx}  
\usepackage{dcolumn}  
\usepackage{bm}  
\usepackage{epstopdf}
\usepackage{times}
\usepackage{mathptmx}
\usepackage{wasysym}

\begin{document}

\title{Donor-driven spin relaxation in multi-valley semiconductors}
\author{Yang~Song}
\affiliation{Department of Electrical and Computer Engineering, University of Rochester, Rochester, New York 14627, USA}

\author{Oleg~Chalaev}
\altaffiliation{Yang~Song and Oleg~Chalaev have equal contribution to this work.}
\affiliation{Department of Electrical and Computer Engineering, University of Rochester, Rochester, New York 14627, USA}

\author{Hanan~Dery}
\altaffiliation{hanan.dery@rochester.edu}
\affiliation{Department of Electrical and Computer Engineering, University of Rochester, Rochester, New York 14627, USA}
\affiliation{Department of Physics and Astronomy, University of Rochester, Rochester, New York 14627, USA}

\begin{abstract}
We present a theory for spin relaxation of electrons due to scattering off the central-cell potential of impurities in silicon. Taking into account the multivalley nature of the conduction band and the violation of translation symmetry, the spin-flip amplitude is dominated by this short-range impurity scattering after which the electron is transferred to a valley on a different axis in \textit{k}-space (the so called \textit{f}-process).  These \textit{f}-processes dominate the spin relaxation at all temperatures, where scattering off the impurity central-cell dominate at low temperatures, and scattering with $\Sigma$-axis phonons at elevated temperatures. To the best of our knowledge, the theory is the first to explain and accurately quantify the empirically-found dependence of spin relaxation on the impurity identity. Accordingly, the new formalism fills a longstanding gap in the spin relaxation theory of \textit{n}-type silicon, and it is valuable for characterization of silicon-based spintronic devices.
\end{abstract}

\maketitle

A major quest in semiconductor spintronics is genuine electrical spin injection from ferromagnetic metals.\cite{Hanbicki_APL03,Zutic_RMP04,Crooker_Science05,Appelbaum_Nature07,Shiraishi_PRB11,Ando_APL11,Dery_APL11,Jansen_NatMater12} In spite of the fabrication challenge, electrical spin injection has no intrinsic limitation that prevents the development of deep-submicron devices. In order to electrically inject spin-polarized currents, however, one cannot use ohmic contacts between semiconductors and metals due the so-called conductivity mismatch problem.\cite{Schmidt_PRB00,Rashba_PRB00,Fert_PRB01} Accordingly, electrical techniques are largely limited to spin injection by ballistic hot electrons,\cite{Appelbaum_Nature07,Li_PRL13,Lu_APL13} or tunneling across thin barriers. The latter approach can be integrated with the mainstream Si technology but requires narrowing down the Schottky depletion region to a few nm in order to enable measurable tunneling of spin-polarized electrons.\cite{Hanbicki_APL03,Crooker_Science05,Shiraishi_PRB11,Ando_APL11} Such tunnel barriers can be fabricated by introducing degenerate \textit{n}-type interfaces which have the detrimental effect of enhanced spin relaxation due to electron-impurity scattering. To date, existing theories neither can quantify nor explain the spin relaxation in heavily doped \textit{n}-type diamond crystal semiconductors. The lack of understanding hinders development of spintronics devices with tailored spin relaxation, thereby hampering the progress of this research field. 

A salient feature of spin relaxation in \textit{n}-type silicon is a strong dependence on the donor atom.\cite{Ue_PRB71,Quirt_PRB72,Pifer_PRB75,Ochiai_PSS76,Zarifis_PRB98} For example, it has been long recognized from electron paramagnetic resonance (EPR) experiments that the spin lifetime is about 100 times shorter in heavily antimony-doped silicon (Si:Sb) than in phosphorus-doped silicon (Si:P) with comparable impurity concentration.\cite{Pifer_PRB75,Zarifis_PRB98} This finding contradicts the traditional Elliott picture for spin relaxation, in which the probability for an electron to flip its spin is governed by the spin-orbit coupling of the host material (Si in this example), whereas the identity of the scattering center is of little importance.\cite{Elliott_PR54,Yafet_SSP63} In addition, the predicted proportionality between mobility and spin relaxation time in the Elliott-Yafet mechanism seems at odds with empirical values in \textit{n}-type Si. That is, the spin relaxation is markedly different in Si:P, Si:As or Si:Sb with comparable impurity concentration,\cite{Ue_PRB71,Quirt_PRB72,Pifer_PRB75,Ochiai_PSS76,Zarifis_PRB98} while the mobility is essentially the same. \cite{Morin_PR54,Wolfstirn_JPCS60,Furukawa_JPSJ61,Granacher_JPCS67,Mousty_JAP74,Ralph_PRB75,Masetti_IEEE83,Grujin_JAP98}

\begin{figure*}[t]
    \includegraphics[width=14.5cm]{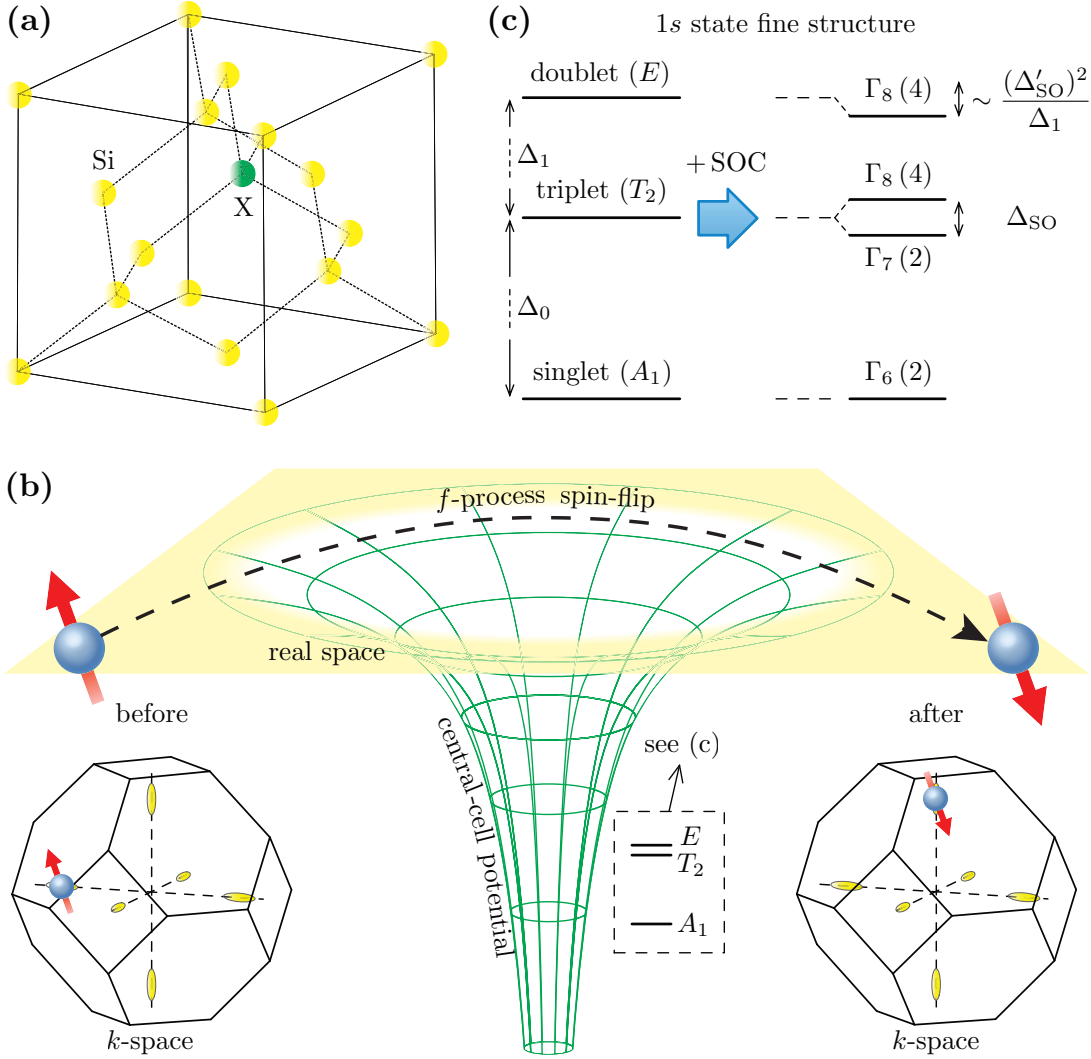}
    \caption{  (a) Substitutional impurity atom in a silicon crystal host. We consider typically employed group V donors such as X = \{P, As, Sb\}.  (b) Scheme of the dominant impurity-driven spin relaxation mechanism. The spin flip is governed by scattering off the central cell potential after which the conduction electron is transferred to a valley on a different crystal axis in \textit{k}-space. (c) Fine structure of the 1\textit{s} state due to the central cell potential. The scattering amplitude in (b) is governed by the impurity spin-orbit coupling parameters $\Delta_{\rm{so}}$ and $\Delta_{\rm{so}}^{'}$.} \label{fig:Si_imp_scheme}
\end{figure*}

We develop a new approach for the problem showing how short-range impurity scattering dominates the spin relaxation. Since the spin-orbit coupling (SOC) is localized within the immediate vicinity of the atomic core, the spin-flip scattering is governed by the difference between the potentials of the impurity and host atoms in the central cell region. Figure~\ref{fig:Si_imp_scheme}(a) shows an example for substitutional impurity atom surrounded by four host atoms in a tetrahedral molecular geometry. These impurities have $T_{\rm{d}}$ point-group symmetry and constitute the vast majority of donors and acceptors in silicon. To unveil the underlying spin relaxation induced by these impurities, we take into account the multivalley nature of the conduction band and the violation of translation symmetry. When considered together, we will show that the spin-flip amplitude is dominated by a short-range impurity scattering after which the electron is transferred to a valley on a different axis in \textit{k}-space (the so called \textit{f}-process). This spin-flip mechanism is schematically shown in Fig.~\ref{fig:Si_imp_scheme}(b). The impurity-induced spin relaxation is evidently weaker for intravalley scattering in which the electron remains in the same valley, or for intervalley \textit{g}-process in which the electron is transferred to the other valley on the same crystallographic axis. The vanishing amplitude of the \textit{g}-process scattering can be shown by invoking time reversal symmetry and of the intravalley scattering by any symmetry operation from the $T_{\rm{d}}$ point-group that includes reflection (Appendix \ref{app:general}). 

Another aspect of the theory relies on the fundamental relation between the scattered and bound states of an impurity potential.\cite{Scattering_Book} This relation allows us to quantify the spin-flip amplitude from the empirically known SOC-induced splitting of the donor-state spectral lines.  The left part of Fig.~\ref{fig:Si_imp_scheme}(c) shows the familiar energy levels of donor states in silicon. Due to the valley-orbit coupling within the central cell, this 1\textit{s} state is split into spin-independent singlet, doublet and triplet states where the overall 6-fold multiplicity comes from the number of conduction edge states (valley centers).\cite{Kohn_SSP57} Taking into account the SOC of the impurity, two important energy scales are relevant. The first one  corresponds to splitting of the triplet state to four-fold ($\Gamma_8$) and two-fold ($\Gamma_7$) spin dependent states, as shown in the right part of Fig.~\ref{fig:Si_imp_scheme}(c). The SOC-induced splitting is known empirically: $\Delta_{\rm{so}}$$\,$$\approx$$\,$0.03~meV for Si:P, 0.1~meV for Si:As, and 0.3~meV for Si:Sb.\cite{Castner_PR67,Aggrawal_PR65} The second energy scale is more subtle and comes from spin-dependent interaction between the four-fold degenerate states [two $\Gamma_8$ levels in Fig.~\ref{fig:Si_imp_scheme}(c)]. This interaction is manifested by a small added contribution, $\Delta_{\rm{so}}^{'}$, to the splitting of these states (Appendix~\ref{app:cc}),
\begin{eqnarray}\label{eq:DeltaSoprime}
\varepsilon_{\Gamma_8} \rightarrow \frac{1}{2}\left[ (\varepsilon_{T_2}+\varepsilon_{E}) \pm \sqrt{ \Delta_1^2 + (\Delta_{\rm{so}}^{'})^2} \right]. \nonumber
\end{eqnarray}
$\varepsilon_{T_2(E)}$ is the spin-independent energy of the triplet (doublet) state where $\Delta_1=\varepsilon_{T_2}-\varepsilon_{E}$  is their valley-orbit induced splitting (typically much larger than $\Delta_{\rm{so}}^{'}$). We note that the presence of $\Delta_{\rm{so}}^{'}$ is mandated by symmetry and that its magnitude should be commensurate with that of $\Delta_{\rm{so}}$ since both originate from similar impurity orbitals (Appendix~\ref{app:cc}). Indeed, we will show that the theory agrees with empirical values of the spin relaxation for $|\eta| \equiv |\Delta_{\rm{so}}^{'}/\Delta_{\rm{so}}| \approx 2$.

\begin{figure*}[t]
\includegraphics[width=14cm]{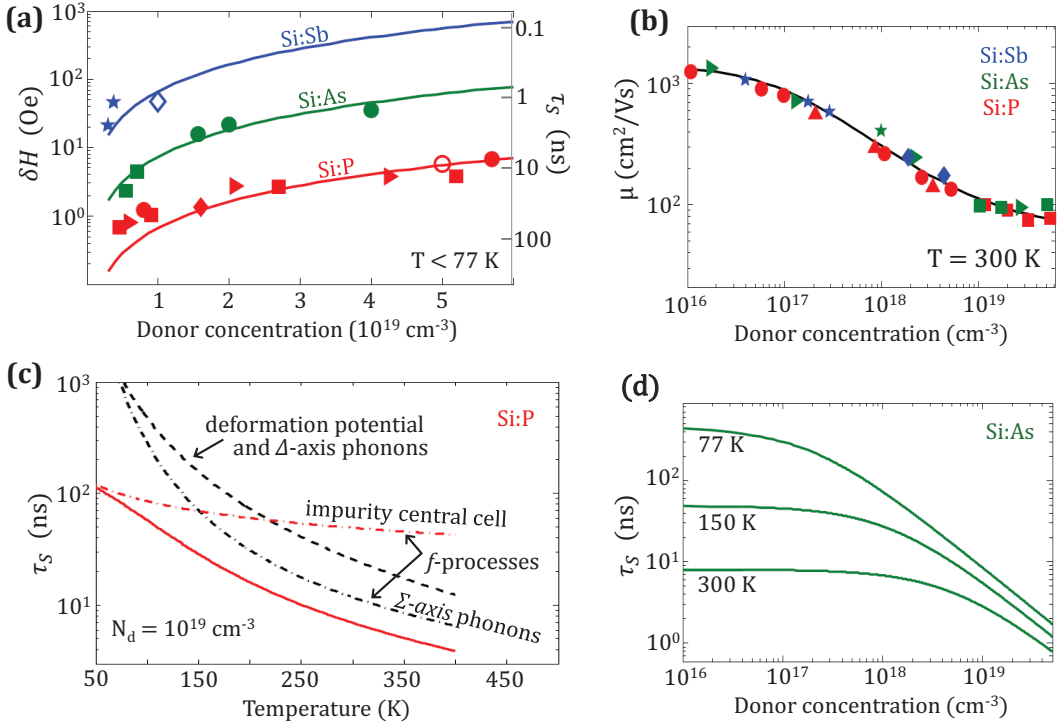}
\caption{(a) Spin relaxation in heavily doped \textit{n}-type silicon for three common donor types: phosphorous (Si:P),  arsenic (Si:As), and antimony (Si:Sb). Solid lines denote the theory results for the average spin lifetime (right axis), and solid symbols denote empirical values of the measured linewidth in EPR and spin injection experiments (left axis, $\Circle$ [\onlinecite{Shiraishi_PRB11}], $\lozenge$ [\onlinecite{Ando_arXiv14}], $\blacklozenge$ [\onlinecite{Ue_PRB71}], $\blacktriangleright$ [\onlinecite{Quirt_PRB72}], $\blacksquare$ [\onlinecite{Pifer_PRB75}], $\CIRCLE$ [\onlinecite{Ochiai_PSS76}], $\bigstar$ [\onlinecite{Zarifis_PRB98}]). The scales on the left and right axes are related by $\delta H= 1/\gamma_e \tau_s$ where $\gamma_e$=1.7$\times$10$^{7}$ ~s$^{-1}$$\cdot$ Oe$^{-1}$ is the electron gyromagnetic ratio.  (b) Room-temperature mobility versus donor concentration showing a marginal dependence on donor identity ($\blacktriangleright$ [\onlinecite{Morin_PR54}], $\bigstar$ [\onlinecite{Wolfstirn_JPCS60}],  $\blacklozenge$ [\onlinecite{Furukawa_JPSJ61}], $\blacktriangle$ [\onlinecite{Granacher_JPCS67}], $\CIRCLE$ [\onlinecite{Mousty_JAP74}], $\blacksquare$ [\onlinecite{Masetti_IEEE83}]). (c) Temperature dependence of $\tau_s$ in 10$^{19}$~cm$^{-3}$ Si:P (solid line). Intervalley \textit{f}-processes dominate the relaxation at all temperatures. (d) Doping concentration dependence of $\tau_s$ in Si:Ar at three temperatures. The decay of $\tau_s$ when entering the metallic regime ($N_d > 2\times10^{18}$) is due to transition from electron-phonon to electron-impurity dominated relaxation. } \label{fig:Si_imp_theory}
\end{figure*}

Using $\Delta_{\rm{so}}$ and $\eta$, we quantify the dominant spin relaxation effect in heavily doped \textit{n}-type Si. It refers to the \textit{f}-process spin flip in which a conduction electron is scattered off the central-cell potential [Fig.~\ref{fig:Si_imp_scheme}(b)]. The corresponding scattering matrix element is  (Appendix \ref{app:general}),
\begin{eqnarray}\label{eq:Usf}
U_{sf}^f = \frac{\pi a_B^3}{V} \left[ \frac{ie^{i\phi}}{6}\sin{\theta} - \frac{\eta(1+i)}{\sqrt{12}}\left( i\cos^2{\tfrac{\theta}{2}}  + \sin^2{\tfrac{\theta}{2}} e^{2i\phi} \right)  \right] \Delta_{\rm{so}}, \nonumber
\end{eqnarray}
where $V$ is the crystal volume and $a_B \approx 2$~nm is the electron Bohr radius in Si . The normalization factor $\pi a_B^3/V$ is due to the fact that $\Delta_{\rm{so}}$ and $\eta$  are bound-state parameters whereas the transition amplitude is that of scattered states that extend across the crystal. The polar and azimuthal angles ($\theta$ \& $\phi$) define the spin orientation, where the polar angle is measured from the normal direction to the plane defined by two valley axes of the \textit{f}-process. To better understand this angular dependence, we consider an example where electrons have a net spin polarization along the \textit{z}-axis. The spin-flip amplitude is calculated by assigning $\theta = 0$ for scattering between $\pm x$ and $\pm y$ valleys since the net spin polarization along the \textit{z}-axis is parallel to the normal of the $xy$~plane. Similarly, we assign $\theta = \pi/2$ \& $\phi = 0(\pi/2)$ for scattering between $\pm z$ and $\pm x(y)$ valleys since the net spin-polarization along the \textit{z}-axis is perpendicular to the normal of the $xz$~($yz$) plane. After averaging over all valley configurations and summing over final states, we get that the spin relaxation of a conduction electron with energy $\varepsilon$ above the band edge is
\begin{eqnarray}\label{eq:tau_s_imp}
\frac{1}{\tau_s(\varepsilon)} =   \frac{2\pi N_d m_e a_B^6}{9\hbar^4 } \sqrt{2m_e \varepsilon}  (4|\eta|^2+1 ) \Delta_{so}^2  . \nonumber
\end{eqnarray}
$N_d$  is the donor concentration and $m_e = 0.32m_0$  is the electron effective mass in Si. In the high-temperature regime, we can assign $\varepsilon \approx K_BT$  and get that the effective spin relaxation rate scales with $\sqrt{T}$. In the opposite limit ($\varepsilon_F \gtrsim K_BT$), the average spin lifetime is temperature-independent and found by assigning $\varepsilon \approx \varepsilon_F$  where $\sqrt{2m_e \varepsilon_F}/\hbar \approx \sqrt[3]{3\pi^2 N_d}$. The solid lines in Fig.~\ref{fig:Si_imp_theory}(a) show the calculated spin lifetime in this limit for $|\eta|=2$. The symbols are compiled results from seven different experiments.\cite{Shiraishi_PRB11,Ue_PRB71,Quirt_PRB72,Pifer_PRB75,Ochiai_PSS76,Zarifis_PRB98,Ando_arXiv14} The theory shows excellent agreement with experiment apart from a small discrepancy for Si:P when approaching the critical metal-to-insulator transition ($\sim 2 - 4 \cdot 10^{18}$~cm$^{-3}$). The relatively long spin lifetime of Si:P in this regime may have another contribution due to remnant effects of the impurity band.\cite{Anderson_JPSJ54} For the case of \textit{n}-type Si:Sb, we could not find experimental results for $N_d > 10^{19}$~cm$^{-3}$ possibly due to strong antimony segregation when growing samples at these doping levels.

Unlike the spin relaxation of conduction electrons, their mobility is not affected by the identity of the donor atom. Figure~\ref{fig:Si_imp_theory}(b) shows a compilation of empirical mobility values along with the theoretical curve.\cite{Masetti_IEEE83} The mobility is governed mostly by the potential tail of ionized impurities away from the central-cell, $U(r)\propto \exp{(-\kappa r)}/r$ where $\kappa^{-1}$ is the screening length. This ionized impurity potential is identical for donors from the same column of the periodic table, explaining why the proportionality factor between mobility and spin relaxation changes dramatically when replacing the substitutional donor. Whereas central cell effects induce marginal corrections for mobility,\cite{Ralph_PRB75,Masetti_IEEE83,Grujin_JAP98} they are indispensable for spin relaxation in multivalley materials.

Figure~\ref{fig:Si_imp_theory}(c) shows the temperature dependence of the spin lifetime in 10$^{19}$~cm$^{-3}$ Si:P (solid line). The calculation considers both electron-impurity and electron-phonon interactions. The latter has already been quantified,\cite{Cheng_PRL10,Restrepo_PRL12} and was shown to be dominated by an intervalley \textit{f}-process due to scattering with shortwave $\Sigma$-axis phonons.\cite{Li_PRL11} Also shown are contributions from electron-phonon intravalley and $g$-process scattering due to interactions with the crystal deformation potential and $\Delta$-axis phonons, respectively.\cite{Li_PRL11} We find that intravalley scattering off the potential tail of ionized impurities, which largely sets the mobility in doped semiconductors, is responsible for orders of magnitudes weaker spin relaxation than all shown  mechanisms in Fig.~\ref{fig:Si_imp_theory}(c) [Appendix~\ref{app:long} and \ref{app:taus}]. The dash-dotted lines in Fig.~\ref{fig:Si_imp_theory}(c) show that the spin relaxation is dominated by \textit{f}-processes where at low temperatures it comes from scattering off the impurity central-cell and at elevated temperatures from scattering with $\Sigma$-axis phonons. These processes dominate the spin relaxation in both electron-phonon and electron-impurity scattering since they do not vanish at the lowest order. Specifically, space inversion and time reversal symmetries affect only the intravalley and $g$-process scattering rendering them weak effects.\cite{Song_PRB12}  Figure~\ref{fig:Si_imp_theory}(d) shows the dependence of spin relaxation on doping concentration in Si:As. The spin relaxation enhancement is evident when entering the metallic regime ($N_d > 2\times10^{18}$) due to a change in the \textit{f}-process dominant mechanism from electron scattering with $\Sigma$-axis phonons to electron scattering off the central-cell potential of impurities.

Ramifications of the studied central-cell-driven spin relaxation extend beyond diamond structure crystals. The prerequisite condition for a strong signature of the effect is that conduction electrons populate distinct \textit{k}-space regions that cannot be connected by time reversal symmetry. This condition is satisfied by most known metals, rock salt crystals, or even oxide heterostructures. It does not apply in materials where thermal electrons populate a single zone-center valley such as in GaAs or if all the distinct valleys are related by time reversal such as in graphene. In these cases, spin flips due to short-range impurity scattering vanish in the lowest order. Of all materials that obey the prerequisite condition, one should focus on crystals that respect space inversion symmetry. In multivalley materials that lack a space inversion center such as AlAs or GaP semiconductors with zinc-blende crystal structure, the Dyakonov-Perel spin relaxation mechanism can compete with the studied spin-flip effect at elevated temperatures.\cite{Dyakonov_JETP73}  In addition, one should also focus on materials whose Fermi surfaces exludes spin hot-spots (regions in \textit{k}-space where band degeneracy is lifted by the SOC). Spin hot-spots appear, for example, in small regions of the Fermi surface of aluminum and can dominate its spin relaxation due to strong spin mixing of states in these regions.\cite{Fabian_PRL98} These spin hot-spots also explain the ultra-short spin lifetime of holes in unstrained bulk $sp^3$ semiconductors with diamond, zincblende and wurtzite crystal structures (e.g., Si, GaAs, and ZnO).\cite{Hilton_PRL02,Loren_PRB11} In these semiconductors, the SOC lifts the three-fold band degeneracy at the top of the valence band, rendering a strong spin mixing of light and split-off holes.

In conclusion, we have identified a general spin relaxation mechanism in multivalley materials that so-far has been overlooked. The new formalism fills a longstanding gap in the spin relaxation theory of \textit{n}-type silicon, and it is valuable for characterization of silicon-based spintronic devices. For example, it can be used to optimize on-chip spin communications over millimeter length scales.\cite{Dery_APL11,Zutic_NatureMater11} Knowing that the intervalley \textit{f}-process dominates the spin relaxation, one can enhance the spin lifetime by lifting the valley degeneracy.\cite{Dery_APL11,Tang_PRB12,Osintsev_AdvMater13}  Application of uniaxial compressive strain along the [$001$] crystallographic direction raises the energies of the $\pm x$ and $\pm y$ valleys while lowers those of the $\pm z$ valleys. As a result, the $\pm x$ and $\pm y$ valleys are depopulated and if their energy splitting from the $\pm z$ valleys is large compared with $K_BT$, then in the lowest order,  electrons experience neither elastic nor inelastic intervalley scattering via impurities or shortwave phonons, respectively. Finally, the analysis can be repeated for \textit{n}-type Ge which has  four valleys centered at the \textit{L}~point of the Brillouin zone. Lifting the valley degeneracy in Ge by strain along the [$111$] crystallographic axis can lead to exceptionally long spin lifetime given the absence of $g$-processes and ultra-weak intravalley spin flips in this material.\cite{Li_PRB12}

This work is supported by NSF and DTRA Contracts No. ECCS-1231570 and HDTRA1-13-1-0013, respectively.

\begin{widetext}

\appendix

\section{General selection rules from symmetry considerations}\label{app:general}

The scattering matrix element between two band states $\psi_1$ and $\psi_2$ due to a substitutional impurity  is generally expressed as
\begin{eqnarray}
U_{12}=\langle \psi_2 |U | \psi_1\rangle.
\end{eqnarray}
$U$ is the difference between the potentials of the impurity and host atoms. Regardless of its details, $U$ is invariant under all operations of $T_d$ point group. Table~\ref{tab:CharactorTable_T_d} shows the character table of $T_d$ group following the notation of Bradley and Cracknell.\cite{Bradley_Cracknell72}
\begin{table}[!htbp]
{\begin{center}
\caption{Character table of $T_d$ point group.}
\label{tab:CharactorTable_T_d}
\renewcommand{\arraystretch}{1.2}
\begin{tabular}{cc|ccc ccc cc}
\hline \hline
\multicolumn{2}{c|}{$T_d$}
&\;$E$\;& \;$\bar{E}$\; & \;$3C_2,3\bar{C}_2$\; & \; $4C^+_3,4C^-_3$ \; & \;$4\bar{C}^+_3,4\bar{C}^-_3$ \;& \;$3S^+_4,3S^-_4$ \;& \;$3\bar{S}^+_4,3\bar{S}^-_4$ \;& \;$6\sigma,6\bar{\sigma}$
\\ \hline
$A_1$ & $\Gamma_1$           & 1 & 1        &1              &1                            & 1                          & 1            &1                           &1
  \\
$A_2$ &$\Gamma_2$         & 1 & 1        &1              &1                            & 1                          & $-1$            &$-1$                           &$-1$
  \\
$E$ &$\Gamma_3$       & 2 & 2        &2              &$-1$                            &$-1$                        &0          &0                         &0
\\
$T_1$ &$\Gamma_4$         & 3 & 3        &$-1$              &0    &0     &1  &1    &$-1$
  \\
$T_2$ &$\Gamma_5$        & 3 & 3        &$-1$                  &0    &0     &$-1$  &$-1$    &$1$
\\
$\bar{E}_1$ &$\Gamma_6$   & 2 & $-2$    &0       &1        &$-1$     &$\sqrt{2}$            &$-\sqrt{2}$ &0
\\
$\bar{E}_2$ &$\Gamma_7$  & 2 & $-2$    &0       &1        &$-1$     &$-\sqrt{2}$            &$\sqrt{2}$ &0
 \\
$\bar{F}$ &$\Gamma_8$      & 4 &$-4$    &0      &$-1$      &1     &0   &0    &0
    \\
\hline\hline
\end{tabular}
\end{center}}
\end{table}

The matrix elements do not vanish when $\psi_2^*(\mathbf{r})\psi_1(\mathbf{r})$ or part of it belong to the identity irreducible representation (IR) $A_1$. So naturally, we first express $\psi_i$ in terms of the IRs of $T_d$ (i.e., symmetrization). For the leading order effect, we now consider $\psi$ only from the conduction band valley minima.

\subsection{Without spin degree of freedom}

$\psi$ consists of six degenerate basis states and they can be symmetrized into the following 6 states according to Table~\ref{tab:CharactorTable_T_d},\cite{Kohn_SSP57}
\begin{subequations}\label{eq:symmetrize-nospin}
\begin{eqnarray}
\psi_{A_1}&=& \frac{1}{\sqrt{6}}(1, 1, 1, 1, 1, 1);  \\
\psi_{E^I}&=&\frac{1}{2}(1, 1, -1, -1, 0, 0),\\
\psi_{E^{I\!I}}&=&\frac{1}{2\sqrt{3}}(1, 1, 1, 1, -2, -2);\\
\psi_{T_2^{I}}&=&\frac{1}{\sqrt{2}}(1, -1, 0, 0, 0, 0),\\
\psi_{T_2^{I\!I}}&=&\frac{1}{\sqrt{2}}( 0, 0,1, -1, 0, 0),\\
\psi_{T_2^{I\!I\!I}}&=&\frac{1}{\sqrt{2}}( 0, 0, 0, 0,1, -1);
\end{eqnarray}
\end{subequations}
belonging respectively to three IRs of $A_1$, $E$ and $T_2$. Note that the $T_1$ IR in Ref.~[\onlinecite{Kohn_SSP57}] is the $T_2$ IR in our notation. The ordering of the 6 components of a state vector is the valley minima of $+x$, $-x$, $+y$, $-y$, $+z$ and $-z$ axis, respectively. Therefore we have three independent scattering constants, one from each of the IR above.\footnote{Each $\psi_2^* \psi_1$ contains an identity IR ($A_1$) when and only when $\psi_{1,2}$ belong to the same IR. Explicitly, $A_1\times A_1=A_1$, $E\times E=A_1+A_2+E$ and $T_2\times T_2=A_1+E+T_1+T_2$.} It follows
\begin{subequations}
\begin{eqnarray}
U_{A_1 A_1}=\langle \psi_{A_1} |U | \psi_{A_1}\rangle &=& C_1,  \\
U_{E^I E^I}=U_{E^{I\!I} E^{I\!I}} &=& C_2,\\
U_{T_2^{I} T_2^{I}}=U_{T_2^{I\!I} T_2^{I\!I}} = U_{T_2^{I\!I\!I} T_2^{I\!I\!I}}&=& C_3,
\end{eqnarray}
\end{subequations}
where all $C_i$'s are real numbers. Other combinations vanish.

One can easily re-categorize the scattering conventionally into,\cite{Ralph_PRB75}
\begin{subequations}
\begin{eqnarray}
U_{intra}&=&\langle \frac{1}{6}(\sqrt{6}\psi_{A_1}+3\psi_{E^I} +\sqrt{3}\psi_{E^{I\!I}}+ 3\sqrt{2}\psi_{T_2^{I}}) |U | \frac{1}{6}(\sqrt{6}\psi_{A_1}+3\psi_{E^I} +\sqrt{3}\psi_{E^{I\!I}}+ 3\sqrt{2}\psi_{T_2^{I}}) \rangle
\nonumber\\
 &=& \frac{1}{6}\left(C_1+2C_2+3C_3\right),  \\
U_{g}&=& \langle \frac{1}{6}(\sqrt{6}\psi_{A_1}+3\psi_{E^I} +\sqrt{3}\psi_{E^{I\!I}}+ 3\sqrt{2}\psi_{T_2^{I}} |U | \frac{1}{6}(\sqrt{6}\psi_{A_1}+3\psi_{E^I} +\sqrt{3}\psi_{E^{I\!I}}- 3\sqrt{2}\psi_{T_2^{I}} \rangle \nonumber\\
 &=& \frac{1}{6}\left(C_1+2C_2-3C_3\right),\\
U_{f}&=& \langle \frac{1}{6}(\sqrt{6}\psi_{A_1}+3\psi_{E^I} +\sqrt{3}\psi_{E^{I\!I}}+ 3\sqrt{2}\psi_{T_2^{I}} |U | \frac{1}{6}(\sqrt{6}\psi_{A_1}-3\psi_{E^I} +\sqrt{3}\psi_{E^{I\!I}}- 3\sqrt{2}\psi_{T_2^{I\!I}} \rangle\nonumber\\
 &=& \frac{1}{6}\left(C_1-C_2\right).
\end{eqnarray}
\end{subequations}

\subsection{With spin degree of freedom}
Now $\psi$ consists of a basis with twelve degenerate states. We can symmetrize them on the basis of spin-independent results in Eq.~(\ref{eq:symmetrize-nospin}). A pure spin transform as $\bar{E}_1$ in Table~\ref{tab:CharactorTable_T_d}. Firstly, from
\begin{eqnarray}\label{eq:single2double}
A_1 \times \bar{E}_1=\bar{E}_1, \quad E \times \bar{E}_1=\bar{F},\quad  T_2 \times \bar{E}_1=\bar{E}_2 + \bar{F},
\end{eqnarray}
one can immediately know that there is no spin flip between $\psi_{A_1}$ or between $\psi_E$,  one independent spin-flip constant between $\psi_{T_2}$, as well as one spin-flip constant coupling $\psi_E$ to $\psi_{T_2}$.\footnote{Each identity comes out from intra-IR coupling, $\bar{E}_1\times \bar{E}_1=A_1+T_1$, $\bar{F}\times \bar{F}=A_1+A_2+E+2T_1+2T_2$ and $\bar{E}_2\times \bar{E}_2=A_1+T_1$.}

The SOC does not lift the spin degeneracy in $A_1$, and the new scattering constant is approximately the one without the spin degree of freedom. One can set the new IR $\bar{E}_1$ of doubled dimension to consist of two basis $\{ A_1\Uparrow, A_1\Downarrow\}$, leading to the following non-vanishing scattering combinations
\begin{subequations}\label{eq:scattering-const-spin}
\begin{eqnarray}
U_{A_1\Uparrow A_1\Uparrow}=U_{A_1\Downarrow A_1\Downarrow}=C'_1\approx C_1.
\end{eqnarray}
Similarly, for the doublet state of the single group ($E$) we get
\begin{eqnarray}
U_{E^{I}\Uparrow E^{I}\Uparrow}=U_{E^{I}\Downarrow E^{I}\Downarrow} =U_{E^{I\!I}\Uparrow E^{I\!I}\Uparrow} =U_{E^{I\!I}\Downarrow E^{I\!I}\Downarrow} =C'_2\approx C_2.
\end{eqnarray}
It is more complicated for the spin-independent triplet state ($T_2$). The SOC splits it into spin-dependent doublet and a quartet. The respective two scattering constants are formally,
\begin{eqnarray}
U_{\bar{E}_2^{I} \bar{E}_2^{I}}= U_{\bar{E}_2^{I\!I} \bar{E}_2^{I\!I}} &=& C_4,\\
U_{\bar{F}^{I} \bar{F}^{I}}= U_{\bar{F}^{I\!I} \bar{F}^{I\!I}} = U_{\bar{F}^{I\!I\!I} \bar{F}^{I\!I\!I}}= U_{\bar{F}^{I\!V} \bar{F}^{I\!V}}&= &C_5.
\end{eqnarray}
\end{subequations}
The main task is to find the correct symmetrized wavefunctions for the two resulting $\bar{E}_2$ and $\bar{F}$ IRs. We find the following basis states, satisfying Table~\ref{tab:CharactorTable_T_d},
\begin{subequations}
\begin{eqnarray}
\psi_{\bar{E}_2^{I}}&=&\frac{1}{\sqrt{6}}(1\Uparrow, -1\Uparrow, -i\Uparrow, i\Uparrow, -1\Downarrow, 1\Downarrow),\\
\psi_{\bar{E}_2^{I\!I}}&=&\frac{1}{\sqrt{6}}(1\Downarrow, -1\Downarrow, i\Downarrow, -i\Downarrow, 1\Uparrow, -1\Uparrow);\\
\psi_{\bar{F}^{I}}&=&\frac{1}{2}(1\Uparrow, -1\Uparrow,i\Uparrow, -i\Uparrow, 0, 0),\\
\psi_{\bar{F}^{I\!I}}&=&\frac{1}{2}(1\Downarrow, -1\Downarrow,-i\Downarrow, i\Downarrow, 0, 0),\\
\psi_{\bar{F}^{I\!I\!I}}&=&-\frac{1}{2\sqrt{3}}(1\Uparrow, -1\Uparrow,-i\Uparrow, i\Uparrow, 2\Downarrow, -2\Downarrow),\\
\psi_{\bar{F}^{I\!V}}&=&-\frac{1}{2\sqrt{3}}(1\Downarrow, -1\Downarrow,i\Downarrow, -i\Downarrow, -2\Uparrow, 2\Uparrow).\\
\end{eqnarray}
\end{subequations}
To be definite, we set the spin orientation in the above basis along the $z$ direction (it can also be along other directions by linear combinations of spin-up along $+z$ and spin-down along $-z$).
We can estimate $C_4\approx C_5 \approx C_3$ if the spin-flip scattering strength is much smaller than the spin-conserving one.\footnote{We can see them explicitly. For example, without SOC, $U_{\bar{E}_2^{I} \bar{E}_2^{I}}$ reduces to $\frac{1}{3}(U_{T^I_2\Uparrow T^I_2\Uparrow}+U_{T^{I\!I}_2\Uparrow T^{I\!I}_2\Uparrow}+U_{T^{I\!I\!I}_2\Downarrow T^{I\!I\!I}_2\Downarrow})$, $U_{\bar{F}^{I} \bar{F}^{I}}$ reduces to $\frac{1}{2}(U_{T^I_2\Uparrow T^I_2\Uparrow}+U_{T^{I\!I}_2\Uparrow T^{I\!I}_2\Uparrow})$, etc.} Lastly, we relate the two sets of 4-dimensional $\bar{F}$ IRs originated from $E$ and $T_2$ [Eq.~(\ref{eq:single2double})]. They can be made to transform exactly the same (a similar representation matrix for every operation). By the basis states defined above, we find
\begin{subequations}\label{eq:Fbar-mapping}
\begin{eqnarray}
\begin{array}{rcl}
\psi_{{E}^{I}\Uparrow}&\leftrightarrow&\psi_{\bar{F}^{I\!V}},\\
\psi_{{E}^{I\!I}\Uparrow}&\leftrightarrow&\psi_{\bar{F}^{I\!I}},\\
\psi_{{E}^{I}\Downarrow}&\leftrightarrow&\psi_{\bar{F}^{I\!I\!I}},\\
\psi_{{E}^{I\!I}\Downarrow}&\leftrightarrow&\psi_{\bar{F}^{I}}.\\
\end{array}
\end{eqnarray}
As a result, the last spin-flip scattering constant is
\begin{eqnarray}
U_{E^{I}\Uparrow \bar{F}^{I\!V}}=U_{{E}^{I\!I}\Uparrow \bar{F}^{I\!I}} =U_{{E}^{I}\Downarrow \bar{F}^{I\!I\!I}} =U_{{E}^{I\!I}\Downarrow \bar{F}^{I}} =C_6.
\end{eqnarray}
Since these are not diagonal matrix elements, additional verification for $C_6$'s is needed from time reversal (TR) operation ($\mathcal{T}$),
\begin{eqnarray}
C_6=\langle \psi_{\bar{F}^{I\!V}}|U|\psi_{E^{I}\Uparrow}\rangle  = \langle\mathcal{T}\psi_{E^{I}\Uparrow}|U| \mathcal{T} \psi_{\bar{F}^{I\!V}}\rangle
=\langle\psi_{E^{I}\Downarrow}|U| \mathcal{T} \psi_{\bar{F}^{I\!I\!I}}\rangle=C^*_6.
\end{eqnarray}
\end{subequations}
So $C_6$ is a real number.

Next we derive the precise spin-dependent scattering, as a function of valley positions and spin orientation, in terms of the above constants $C'_1, C'_2, C_4, C_5$ and $C_6$. First we fix the spin orientation along $z$ direction, and then we show the results for arbitrary spin orientation directions. To do so, we express the states at a single valley edge in terms of the symmetrized ones above,
\begin{subequations}\label{eq:state-per-valley-spin}
\begin{eqnarray}
\psi_{\pm z\Uparrow}&=& \frac{1}{6}\left( \sqrt{6}\psi_{A_1 \Uparrow}-2\sqrt{3} \psi_{E^{I\!I}\Uparrow} \pm (\sqrt{6} \psi_{\bar{E}_2^{I\!I}} +2\sqrt{3}\psi_{\bar{F}^{I\!V}}) \right), \\
\psi_{\pm z\Downarrow}&=& \frac{1}{6}\left( \sqrt{6}\psi_{A_1 \Downarrow}-2\sqrt{3} \psi_{E^{I\!I}\Downarrow} \pm( - \sqrt{6} \psi_{\bar{E}_2^{I}} -2\sqrt{3}\psi_{\bar{F}^{I\!I\!I}}) \right), \\
\psi_{\pm x\Uparrow}&= &\frac{1}{6}\left( \sqrt{6}\psi_{A_1 \Uparrow}+3\psi_{E^{I}\Uparrow}+\sqrt{3} \psi_{E^{I\!I}\Uparrow} \pm( \sqrt{6} \psi_{\bar{E}_2^{I}} +3\psi_{\bar{F}^{I}} -\sqrt{3}\psi_{\bar{F}^{I\!I\!I}} ) \right), \\
\psi_{\pm x\Downarrow}&= &\frac{1}{6}\left( \sqrt{6}\psi_{A_1 \Downarrow}+3\psi_{E^{I}\Downarrow}+\sqrt{3} \psi_{E^{I\!I}\Downarrow} \pm( \sqrt{6} \psi_{\bar{E}_2^{I\!I}} +3\psi_{\bar{F}^{I\!I}} -\sqrt{3}\psi_{\bar{F}^{I\!V}} ) \right),\\
\psi_{+y\Uparrow}&= &\frac{1}{6}\left( \sqrt{6}\psi_{A_1 \Uparrow}-3\psi_{E^{I}\Uparrow}+\sqrt{3} \psi_{E^{I\!I}\Uparrow} + i(\sqrt{6} \psi_{\bar{E}_2^{I}} -3\psi_{\bar{F}^{I}} -\sqrt{3}\psi_{\bar{F}^{I\!I\!I}}) \right), \\
\psi_{+y\Downarrow}&= &\frac{1}{6}\left( \sqrt{6}\psi_{A_1 \Downarrow}-3\psi_{E^{I}\Downarrow}+\sqrt{3} \psi_{E^{I\!I}\Downarrow} -i( \sqrt{6} \psi_{\bar{E}_2^{I\!I}} -3\psi_{\bar{F}^{I\!I}} -\sqrt{3}\psi_{\bar{F}^{I\!V}}) \right)
\end{eqnarray}
\end{subequations}
Together with Eqs.~(\ref{eq:state-per-valley-spin}) and (\ref{eq:scattering-const-spin}),  we readily get for intravalley scattering that
\begin{subequations}\label{eq:intra-g-f-select-rule}
\begin{eqnarray}
U_{+z\Uparrow +z\Uparrow}\stackrel{\sigma_{x y}}{=} U_{+z\Downarrow +z\Downarrow}&=& \frac{1}{6}(C'_1+2C'_2+C_4+2C_5) \approx \frac{1}{6}\left(C_1+2C_2+3{C_3}\right),\\
U_{+z\Uparrow +z\Downarrow}&=& 0; \textrm{(can be verified by, e.g., reflection around (110) plane).}\label{eq:intra_sf_0}
\end{eqnarray}
And for $g$-process scattering we get that
\begin{eqnarray}
U_{+z\Uparrow -z\Uparrow}\stackrel{\sigma_{x y}}{=} U_{+z\Downarrow -z\Downarrow}&=& \frac{1}{6}(C'_1+2C'_2-C_4-2C_5) \approx \frac{1}{6}\left(C_1+2C_2-3{C_3}\right),\\
U_{+z\Uparrow -z\Downarrow}\stackrel{\sigma_{x y}}{=}-i U_{+z\Downarrow -z\Uparrow}&=& 0; \textrm{(can be verified by time reversal symmetry operation).}
\end{eqnarray}
And for $f$-process scattering we get that
\begin{eqnarray}
U_{+x\Uparrow +y\Uparrow}\stackrel{\sigma_{x y}}{=} (U_{+x\Downarrow +y\Downarrow})^*&=& \frac{1}{6}(C'_1-C'_2-iC_4+iC_5) \approx \frac{1}{6}\left(C_1-C_2\right),\\
U_{+x\Uparrow +y\Downarrow}\stackrel{\sigma_{-x y}+TR}{=}-iU_{+x\Downarrow +y\Uparrow}&=& \frac{\sqrt{3}}{6}(1-i)C_6. \label{eq:C6}
\end{eqnarray}
\end{subequations}
The spin flip for arbitrary spin orientation $\mathbf{s}$ ($\theta, \phi$ as the usual angle indexes) is
\begin{eqnarray}\label{eq:U_s}
U_{+x\Uparrow_{\mathbf{s}} +y\Downarrow_{\mathbf{s}}}&=& \langle -\sin\frac{\theta}{2}e^{-i\phi} \psi_{+y\Uparrow} + \cos\frac{\theta}{2} \psi_{+y\Downarrow} |U| \cos\frac{\theta}{2} \psi_{+x\Uparrow} +\sin\frac{\theta}{2}e^{i\phi} \psi_{+x\Downarrow}  \rangle \nonumber\\
&=&  \frac{1}{2}\sin\theta e^{i\phi}(- U_{+x\Uparrow +y\Uparrow} +  U_{+x\Downarrow +y\Downarrow})
+\cos^2\frac{\theta}{2}U_{+x\Uparrow +y\Downarrow}-\sin^2\frac{\theta}{2} e^{2i\phi}U_{+x\Downarrow +y\Uparrow} \nonumber\\
&=& \frac{1}{6}\sin\theta e^{i\phi}i(C_4-C_5)
-\frac{1}{\sqrt{6}}e^{i\pi /4}(i\cos^2\frac{\theta}{2}+\sin^2\frac{\theta}{2} e^{2i\phi})C_6.
\end{eqnarray}
As mentioned in the main text, the magnitude of the spin-flip matrix elements between other valley pairs can be reached by simple rotations of $|U_{+x\Uparrow_{\mathbf{s}} +y\Downarrow_{\mathbf{s}}}|$.

\section{Scattering off the central-cell potential} \label{app:cc}
To calculate or estimate the above spin-flip scattering constants, we could start from the viewpoint of the Elliott-Yafet mechanism. That is, separating the contribution from the spin mixing of the host conduction states (Elliott part), and from the difference between the SOC of the host and that of the impurity (Yafet part). Different emphases have been given on the Elliott part,\cite{Chazalviel_JPCS75} or on the Yafet part.\cite{Castner_PR67,Zarifis_PRB87} One can roughly think of the Elliott part as being dependable on the volume difference of the impurity ion and host ion, while the Yafet part depends on the difference of their SOC.

We briefly show how to write down $C_4-C_5$ and $C_6$ explicitly in terms of spinless wavefunctions $\psi_{\Delta}$'s, perturbations $U_0$ and SOC operators. We can use, e.g., Eqs.~(\ref{eq:U_s}) to achieve them.  The conduction wavefunctions $\phi_i$ at the valley minimum $\mathbf{k}_0$ belong to $\Delta_6$ 2D IR of the $\Delta$ group in diamond structure. In this group, operators $\mathbf{R} \equiv (\hbar/4m^2_0c^2)\bm\nabla V\times \mathbf{p}$ belong to $R_z\sim \Delta'_1$ and $\{R_x, R_y\}\sim \Delta_5$ for the $z$ valley. Wavefunctions (spinless) at the nearby valence band of the same $\mathbf{k}_0$ point also belong to $\Delta_5$. Therefore, in the leading order
\begin{eqnarray}\label{eq:psi_c}
\psi_{z\Uparrow(\Downarrow)_z}=\psi_{\Delta_6\Uparrow(\Downarrow)_z} \approx
\psi_{\Delta_1\uparrow(\downarrow)} + \frac{i\Upsilon_{so}}{E_g}\left(\psi_{\Delta^{(x)}_5\downarrow(\uparrow)}\pm i  \psi_{\Delta^{(y)}_5\downarrow(\uparrow)}\right),
\end{eqnarray}
where $i\Upsilon_{so}=\langle \psi_{\Delta^{(x)}_5}|R_x|\psi_1\rangle = \langle \psi_{\Delta^{(y)}_5}|R_y|\psi_1\rangle$  and $\psi_{\Delta^{(x,y)}_5}$ are the basis functions of the valence band. Similarly, we have
\begin{eqnarray}
\psi_{x\Uparrow(\Downarrow)_s} &\approx&
\psi_{x\Delta_1\uparrow(\downarrow)_z} \pm \frac{i\Upsilon_{so}}{E_g}\left[\pm i(\cos^2\frac{\theta}{2}+\sin^2\frac{\theta}{2} e^{\pm 2i\phi})\psi_{x\Delta^{(y)}_5\downarrow(\uparrow)_z}-\sin\theta e^{\pm i\phi}  \psi_{x\Delta^{(z)}_5\downarrow(\uparrow)_z}\right],
\\
\psi_{y\Uparrow(\Downarrow)_s} &\approx&
\psi_{y\Delta_1\uparrow(\downarrow)_z} \pm \frac{i\Upsilon_{so}}{E_g}\left[-\sin\theta e^{\pm i\phi} \psi_{y\Delta^{(z)}_5\downarrow(\uparrow)_z}  + ( \cos^2\frac{\theta}{2}-\sin^2\frac{\theta}{2} e^{\pm 2i\phi}) \psi_{y\Delta^{(x)}_5\downarrow(\uparrow)_z}\right].
\end{eqnarray}
When $\theta=0$, we have
\begin{eqnarray} \label{eq:C6_integral}
C_6 &=& \sqrt{3}(1+i) U_{+x\Uparrow_z +y\Downarrow_z}\nonumber\\
&=& \sqrt{3}(1+i)\left[
\frac{i\Upsilon_{so}}{E_g} (-\langle  \psi_{y\Delta^{(x)}_5} |U_0| \psi_{x\Delta_1}  \rangle
+ i \langle \psi_{y\Delta_1} |U_0| \psi_{x\Delta^{(y)}_5}\rangle)
+\langle  \psi_{y\Delta_1} |(\bm\nabla U_0\times \mathbf{p})_+|\psi_{x\Delta_1} \rangle \right].
\end{eqnarray}
and when $\theta=\pi/2$, $\phi=-\pi/4$, we have
\begin{eqnarray} \label{eq:C45_integral}
C_4-C_5 &=& 3\sqrt{2}(1-i) U_{+x\Uparrow_s +y\Downarrow_s}\nonumber\\
&=& 3\sqrt{2} \bigg[
\frac{i\Upsilon_{so}}{E_g} ( \langle  ( i\sqrt{2}\psi_{y\Delta^{(z)}_5}  -  \psi_{y\Delta^{(x)}_5} )|U_0| \psi_{x\Delta_1}  \rangle
+  \langle \psi_{y\Delta_1} |U_0|( \psi_{x\Delta^{(y)}_5}+i  \sqrt{2}\psi_{x\Delta^{(z)}_5})\rangle) \nonumber
\\
&&\qquad+\langle  \psi_{y\Delta_1} |(\bm\nabla U_0\times \mathbf{p})_x + (\bm\nabla U_0\times \mathbf{p})_y|\psi_{x\Delta_1} \rangle \bigg].
\end{eqnarray}
In general, $C_4-C_5$ and $C_6$ should be of the same order of magnitude.

More completely and empirically, we can relate and estimate the scattering constants from the experimental measured energy levels of impurities localized states.\cite{Castner_PR67} Importantly the spin splitting  $\Delta_{so}$ of the $T_2$ states is approximately
\begin{eqnarray}
\Delta_{so}&=&\textrm{ 0.03 meV, Si:P; \quad 0.1 meV, Si:As;\quad  0.3 meV, Si:Sb; \quad 1.0 meV, Si:Bi.}
\end{eqnarray}

Next we show the relation between our spin-flip scattering constants $\{C_4-C_5,\,C_6\}$ and $\Delta_{so}$. In order to do so, we first relate the spin-independent constants $C_1,  C_2, C_3$ and the valley-orbit splitting due to the tetrahedral impurity potential. From Kohn's work,\cite{Kohn_SSP57} it is well known that the 6-fold degenerate solution of the effective mass approximation (EMA) is split into 3 states according to the $T_d$ group, just as we did for the scattering problem above. We denote the localized states as $\psi^{loc}$, and the diagonalized Hamiltonian including the tetrahedral impurity potential part is
\begin{eqnarray}
H^{loc}\doteq \left(
\begin{array}{cccccc}
\mathcal{E}_{A_1}&&&&&\\
& \mathcal{E}_{T_2} &&&0&\\
&& \mathcal{E}_{T_2} &&&\\
&&& \mathcal{E}_{T_2} &&\\
&0&&& \mathcal{E}_{E} &\\
&&&&& \mathcal{E}_{E} \\
\end{array}\right),
\end{eqnarray}
where the basis states are $\psi^{loc}_{A_1},\psi^{loc}_{T^I_2},\psi^{loc}_{T^{I\!I}_2},\psi^{loc}_{T^{I\!I\!I}_2}, \psi^{loc}_{E^I},\psi^{loc}_{E^{I\!I}}$, respectively. The energy differences $\mathcal{E}_{A_1}-\mathcal{E}_{E}$ and $\mathcal{E}_{A_1}-\mathcal{E}_{T_2}$ are due to the tetrahedral part of the potential difference between the impurity and host ions. This part apparently mostly comes from the core of the impurity ion. So it has the same origin as that of the $C_1-C_2$ and $C_1-C_3$ obtained above. For the small region of ion core, the smooth EMA (1s) wavefunctions are basically the same as the valley edge conduction wavefunctions, differ only by the volume normalization. So we have
 \begin{eqnarray}
\mathcal{E}_{A_1}-\mathcal{E}_{E}&\equiv& \langle \psi^{loc}_{A_1}| H^{loc}| \psi^{loc}_{A_1}\rangle- \langle \psi^{loc}_{E^I}| H^{loc}| \psi^{loc}_{E^I}\rangle \nonumber\\
&=&
 \langle \psi^{loc}_{A_1}| U| \psi^{loc}_{A_1}\rangle- \langle \psi^{loc}_{E^I}| H^{loc}| \psi^{loc}_{E^I}\rangle \nonumber\\
 &\approx&
\bigg( \langle \psi_{A_1}| U| \psi_{A_1}\rangle- \langle \psi_{E^I}| H^{loc}| \psi_{E^I}\rangle \bigg) \frac{V}{\pi a_{B}^{3}} \nonumber\\
 &=&
\big( C_1-C_2\big) \frac{V}{\pi a_{B}^{3}} ,
\end{eqnarray}
where $V$ is the material volume and $a_B=\hbar^2 \epsilon/m_ee^2$ is the effective Bohr radius ($\sim$ 20 {\AA} in Si). The volume $V$ will be canceled out when we will derive the spin relaxation rate by integrating over final states and considering the density of impurities. Similarly, $C_1-C_3\approx (\mathcal{E}_{A_1}-\mathcal{E}_{T_2})\pi a_{B}^{3}/V$.

Going to the case with SOC, the localized states have 4 energy levels according to the same symmetry arguments we used for the scattering problem.  The $H^{loc}$ is $12\times 12$ and breaks into 4 blocks. The block with basis states $\psi^{loc}_{A_1\Uparrow}, \psi^{loc}_{A_1\Downarrow}$, and the block with basis states $\psi^{loc}_{\bar{E}^I_2}, \psi^{loc}_{\bar{E}^{I\!I}_2}$ are diagonal with eigenvalues $\mathcal{E}'_{A_1}$ and $\mathcal{E}_{\bar{E}_2}$ respectively. Other two blocks with basis states $\psi^{loc}_{E^{I}\Uparrow}, \psi^{loc}_{E^{I}\Downarrow}, \psi^{loc}_{E^{I\!I}\Uparrow}, \psi^{loc}_{E^{I\!I}\Downarrow}$ and $\psi^{loc}_{\bar{F}^{I}}, \psi^{loc}_{\bar{F}^{I\!I}}, \psi^{loc}_{\bar{F}^{I\!I\!I}}, \psi^{loc}_{\bar{F}^{I\!V}}$ are diagonal with eigenvalues $\mathcal{E}'_{E}$ and $\mathcal{E}_{\bar{F}}$, but their cross blocks are also diagonal with elements $\Delta_{so}^{'}$. We focus on this mixed two blocks and diagonalize them. To see it more clearly, we order the basis such that
\begin{eqnarray}
H^{loc}_{\rm partial}\doteq \left(
\begin{array}{cccccccc}
\mathcal{E}'_{E}&\Delta_{so}^{'}&&&&&&\\
\Delta_{so}^{'}& \mathcal{E}_{\bar{F}} &&&&&&\\
&& \mathcal{E}'_{E}&\Delta_{so}^{'} &&&0&\\
&&\Delta_{so}^{'}& \mathcal{E}_{\bar{F}}&& &&\\
&&&& \mathcal{E}'_{E}&\Delta_{so}^{'} &&\\
&0&&&\Delta_{so}^{'}& \mathcal{E}_{\bar{F}}&& \\
&&&&&&\mathcal{E}'_{E}&\Delta_{so}^{'}\\
&&&&&&\Delta_{so}^{'}& \mathcal{E}_{\bar{F}}
\end{array}\right),
\end{eqnarray}
where the basis states are $\psi^{loc}_{E^I\Uparrow},\psi^{loc}_{\bar{F}^I} ,\psi^{loc}_{E^I\Downarrow}, \psi^{loc}_{\bar{F}^{I\!I}},  \psi^{loc}_{E^{I\!I}\Uparrow}, \psi^{loc}_{\bar{F}^{I\!I\!I}} , \psi^{loc}_{E^{I\!I}\Downarrow}, \psi^{loc}_{\bar{F}^{I\!V}} $ with the same symmetry relations as in Eqs.~(\ref{eq:Fbar-mapping}). This part of the Hamiltonian can be diagonalized with eigenvalues
\begin{eqnarray}
\frac{1}{2} \left[\mathcal{E}'_{E}+ \mathcal{E}_{\bar{F}} \pm \sqrt{(\mathcal{E}'_{E}-\mathcal{E}_{\bar{F}})^2  +4(\Delta_{so}^{'})^2} \right] \approx
\{\mathcal{E}'_{E}, \mathcal{E}_{\bar{F}}\},
\end{eqnarray}
with their order undetermined (may be decided by simple argument analogous for $\mathcal{E}_{A_1}, \mathcal{E}_{T_2}$ order)
We argue that $|\mathcal{E}'_{E}-\mathcal{E}_{\bar{F}}|\gg |\Delta_{so}^{'}|$ used for the above approximation. Firstly from experiments one often sees the further splitting of $T_2$ levels is much smaller than the $T_2$ and $E$ level distance. In these situations it has to be $|\mathcal{E}_{E}-\mathcal{E}_{\bar{F}}|\gg |\Delta_{so}^{'}|$. Secondly we see it from the physical origin of $\Delta_{so}^{'}$ and $\mathcal{E}_{E}-\mathcal{E}_{\bar{F}}$: $\Delta_{so}^{'}$ is the coupling of  $\psi_{E}$ and $\psi_{T_2}$ by the {\it SOC} part of the tetrahedral difference (and it is further reduced due to the \textit{different orbital parts} of $\psi_{E}$ and $\psi_{T_2}$, unlike $\mathcal{E}_{\bar{E}_2}-\mathcal{E}_{\bar{F}}$ from the same orbital part of $\psi_{T_2}$); $\mathcal{E}_{E}-\mathcal{E}_{\bar{F}}$ is roughly the valley-orbit splitting due to the spinless tetrahedral potential difference on the $\psi_{E}$ and $\psi_{T_2}$ states.

Thus the localized energy levels correspond to $\mathcal{E}'_{A_1}, \mathcal{E}'_{E}, \mathcal{E}_{\bar{E}_2}, \mathcal{E}_{\bar{F}}$. Using the same arguments we invoked in the spinless case, we get
\begin{eqnarray}
| C_4-C_5|
 &\approx&
 |\mathcal{E}_{\bar{E}_2}-\mathcal{E}_{\bar{F}}| \frac{\pi a_{B}^{3}}{V}=\Delta_{so}\frac{\pi a_{B}^{3}}{V},\\
|C_6| &\approx& |\delta| \frac{\pi a_{B}^{3}}{V}= \Delta'_{so}\frac{\pi a_{B}^{3}}{V}.
\end{eqnarray}
As shown by Eqs.~(\ref{eq:C6_integral}) and ~(\ref{eq:C45_integral}), $\Delta_{so}$ and $\Delta'_{so}$ are of the same order of magnitude.

\section{Scattering off ionized impurities (potential tail outside the central cell region)} \label{app:long}
In this region, the `Elliott' contribution dominates the `Yafet' contribution in general, since the SOC of the Yafet part mostly stems from the rapid change in the potential of central cell correction. The SOC of Elliott part stems from the host and not from the central cell region. We only need to focus on the Elliott part.

In this region, we need to consider intravalley scattering in addition to intervalley scattering, despite the general zero-order selection rule in Eqs.~(\ref{eq:intra-g-f-select-rule}). The reason lies in the suppressed intervalley scattering matrix element due to the Coulomb force. The impurity ($+e$ donor) perturbation outside the central cell can be expressed as (neglecting the tetrahedral wrapping)
\begin{eqnarray}
U(\mathbf{r}) = -\frac{e^2}{4\pi\varepsilon r} e^{-r/\lambda}, \qquad U(\mathbf{q}) = -\frac{e^2}{V\varepsilon (1/\lambda^2+q^2)} .
\end{eqnarray}
where $\lambda$ is the screening length.

We first consider the intravalley scattering. The brief analysis below follows the more systematic effective mass approximation. The spin conserving scattering in $+z$ valley is simply
\begin{eqnarray}
U_{\mathbf{k}_1\Uparrow \mathbf{k}_2\Uparrow} = U^*_{\mathbf{k}_1\Downarrow\mathbf{k}_2\Downarrow}  \approx
\int d \mathbf{r} \psi^*_{\Delta_1} U(\mathbf{r}) \psi_{\Delta_1} e^{i\mathbf{q}\cdot \mathbf{r}}
=  \sum_{n=0}^\infty B_n U(\mathbf{q}+\mathbf{K}_n)
\approx U(\mathbf{q}),
\end{eqnarray}
where $\psi_{\Delta_1}$ is the one in Eq.~(\ref{eq:psi_c}). The spin flip scattering is
\begin{eqnarray}\label{eq:U_flip}
U_{\mathbf{k}_1\Uparrow \mathbf{k}_2\Downarrow} =- U^*_{\mathbf{k}_1\Downarrow\mathbf{k}_2\Uparrow} =
\int d \mathbf{r} \psi^*_{\mathbf{k}_2 \Downarrow} U(\mathbf{r}) \psi_{\mathbf{k}_1\Uparrow}.
\end{eqnarray}
Since from Eq.~(\ref{eq:intra_sf_0}) it is clear the spin-flip amplitude at the valley edge vanishes, we need to expand the state at $\mathbf{k}=\mathbf{k}_0+\mathbf{k}'$ around $\mathbf{k}_0$. We extend the analysis of the $\Delta$ group used in Eq.~(\ref{eq:psi_c}). It is similar to the analysis in Ref.~[\onlinecite{Song_PRB12}] but more compact. As $z\sim \Delta_1$, $R_z\sim \Delta'_1$, $\{R_x, R_y\}\sim \Delta_5$ as well as $\{x,y\}\sim \Delta_5$, the spin-dependent $\mathbf{k}\cdot \mathbf{p}$ method gives
\begin{eqnarray}\label{eq:psi_c_k_up}
\psi_{\mathbf{k} \Uparrow} \approx
e^{i\mathbf{k}' \cdot \mathbf{r}}\left\{\left [\psi_{\Delta_1}+\frac{P}{E_g}(k'_x \psi_{\Delta^{(x)}_5} + k'_y \psi_{\Delta^{(y)}_5})\right]\uparrow
 +\left[-\frac{i P\Upsilon_{so} (k'_x+i k'_y)}{ E^2_g} \psi_{\Delta_1} +\frac{i\Upsilon_{so}}{E_g} (\psi_{\Delta^{(x)}_5} + i  \psi_{\Delta^{(y)}_5}) \right]\downarrow\right\}
\end{eqnarray}
where we used the freedom to fix phases $\hat{T}\hat{S}\psi_{\Delta_i}=\psi_{\Delta_i}$, $i=1,5$, so that
$P=\langle \psi_{\Delta^{(x,y)}_5}|\hbar p_{x,y}/m_0|\psi_1\rangle\approx 10$ $\rm{eV}\cdot {\AA}$ and $\Upsilon_{so} =-i\langle \psi_{\Delta^{(x,y)}_5}|(\hbar/4m^2_0c^2)(\bm\nabla V\times \mathbf{p})_{x,y}|\psi_1\rangle \approx 4.1$ meV are real. $E_g \simeq 4$~eV is the gap between the conduction and valence bands at $\mathbf{k}_0$. Note that $\psi_{\mathbf{k} \Downarrow}= \hat{T}\hat{S} \psi_{\mathbf{k} \Uparrow} $. From Eqs.~(\ref{eq:U_flip}) and (\ref{eq:psi_c_k_up}) we have the biggest contribution to the Elliott process taken from the coefficient product of the same $\Delta$ IR and first component $\mathbf{q}+\mathbf{K}_n=\mathbf{q}$,
\begin{eqnarray}\label{eq:intra_Coub_sf_1st}
U_{\mathbf{k}_1\Uparrow \mathbf{k}_2\Downarrow}  \approx
\frac{2i P\Upsilon_{so} (q_x+iq_y)}{ E^2_g}
 U(\mathbf{q}).
\end{eqnarray}
The next order contribution is from the bigger coefficient product between $\psi_{\Delta_1}$ and $\psi_{\Delta_5}$, by taking the next component $\mathbf{q}+\mathbf{K}_1$'s,
\begin{eqnarray}\label{eq:intra_Coub_sf_2nd}
-\frac{\Upsilon_{so}}{ E_g} \sum_{|\mathbf{K}_n|=|\mathbf{K}_1|} [B^{15x}_n -B^{51x}_n +  B^{15y}_n +B^{51y}_n]U(\mathbf{q}+\mathbf{K}_n),
\end{eqnarray}
where
\begin{eqnarray}
B^{15x}_n = \int d\mathbf{r} \left[\psi^*_{\Delta^{(x)}_5}\psi_{\Delta_1} e^{i \mathbf{K}_n\cdot\mathbf{r}}\right]
\end{eqnarray}
and other $B$'s may be the order of 1. Since $U(\mathbf{q})\propto 1/q^2$ for $q\gg 1/\lambda$ [Eq.~(\ref{eq:U_q})], Eq.~(\ref{eq:intra_Coub_sf_1st}) dominates Eq.~(\ref{eq:intra_Coub_sf_2nd}).

Next we briefly show that although an $f$-process spin flip can happen between valley minima, its magnitude is much smaller than Eq.~(\ref{eq:intra_Coub_sf_1st}) for $q\sim \sqrt{2m_ek_B T}/\hbar$ and is similar to Eq.~(\ref{eq:intra_Coub_sf_2nd}).
\begin{eqnarray}
U_{+x\Uparrow +Y\Downarrow}
&\approx &
\int d \mathbf{r} (\psi_{y\Delta_1}\downarrow-i\frac{\Upsilon_{so}}{E_g} \psi_{y\Delta^{(x)}_5} \uparrow)^* U(\mathbf{r}) (\psi_{x\Delta_1}\uparrow -i\frac{\Upsilon_{so}}{E_g} \psi_{x\Delta^{(y)}_5}\downarrow)\nonumber\\
&\approx& i\frac{\Upsilon_{so}}{E_g}
\sum_\pm (B^{1x5y}_{\pm}- B^{5x1y}_{\pm}) U(\mathbf{k}_{y,0}- \mathbf{k}_{x,0}+(-1,1,\pm 1)\tfrac{2\pi}{a}),
\end{eqnarray}
where
\begin{eqnarray}
B^{1x5y}_\pm = \int d\mathbf{r} \left[u^*_{\Delta^{(x)}_5} u_{\Delta_1} e^{i (-1,1,\pm 1)2\pi/a\cdot\mathbf{r}}\right]
\end{eqnarray}
and other $B$'s could be the order of 1. $u$ is the Bloch part of the wavefunction $\psi$

Now simply extend Eq.~(\ref{eq:intra_Coub_sf_1st}) to arbitrary spin orientation case using Eq.~(\ref{eq:U_s}),
\begin{eqnarray}\label{eq:U_s_intra}
U_{\mathbf{k}_1\Uparrow_{\mathbf{s}} \mathbf{k}_2\Downarrow_{\mathbf{s}}}
&=& \langle -\sin\frac{\theta}{2}e^{-i\phi} \psi_{\mathbf{k}_2\Uparrow} + \cos\frac{\theta}{2} \psi_{\mathbf{k}_2 \Downarrow} |U| \cos\frac{\theta}{2} \psi_{\mathbf{k}_1 \Uparrow} +\sin\frac{\theta}{2}e^{i\phi} \psi_{\mathbf{k}_1 \Downarrow}  \rangle \nonumber\\
&=& \cos^2\frac{\theta}{2}U_{\mathbf{k}_1\Uparrow \mathbf{k}_2\Downarrow}-\sin^2\frac{\theta}{2} e^{2i\phi}U_{\mathbf{k}_1\Downarrow \mathbf{k}_2\Uparrow} \nonumber\\
&=& \frac{2 iP\Upsilon_{so} }{ E^2_g}
 U(\mathbf{q}) \left[\cos^2\frac{\theta}{2} (q_x+iq_y)- \sin^2\frac{\theta}{2} e^{2i\phi} (q_x-iq_y)\right]
\end{eqnarray}

\section{Spin relaxation and its manipulation}\label{app:taus}
Taking the leading order effect, the spin relaxation of a conduction electron with energy $\varepsilon$ above the band edge is
\begin{eqnarray}
\frac{1}{\tau_{s}(\varepsilon)}=\frac{4\pi}{\hbar} \frac{N_dV^2}{(2\pi)^3}
\int d^3k_2 |U_{\mathbf{k}_1\Uparrow \mathbf{k}_2\Downarrow}(\mathbf{s})|^2
\delta(E_{\mathbf{k}_2}-\varepsilon),
 \label{eq:tau_s}
\end{eqnarray}
where $N_d$ is the density of impurities. For scattering off the central-cell impurity potential we assign 
\begin{eqnarray}
|U_{\mathbf{k}_1\Uparrow \mathbf{k}_2\Downarrow}(\mathbf{s})| & \rightarrow &
 \big|\frac{1}{6}\sin\theta e^{i\phi}i\Delta_{so}
-\frac{1}{\sqrt{6}}e^{i\pi /4}(i\cos^2\frac{\theta}{2}+\sin^2\frac{\theta}{2} e^{2i\phi})\Delta'_{so}\big|\frac{\pi a_{B}^{3}}{V}
\end{eqnarray} 
and after integration we get the concise expression for the relaxation rate as shown in the main text (and plotted in Fig. 2). For Coulomb scattering off the potential tail of ionized impurities we assign
\begin{eqnarray} \label{eq:U_tail_long}
|U_{\mathbf{k}_1\Uparrow \mathbf{k}_2\Downarrow}(\mathbf{s})| & \rightarrow & \frac{\hbar^2}{m_e a_B}\frac{2P\Upsilon_{so} }{ V E^2_g} \frac{1}{ 1/\lambda^2+q^2}
\sqrt{(1-\frac{1}{2}\sin^2\theta)  (q^2_x+q^2_y)+\sin^2\theta \left[\frac{1}{2} \cos 2\phi (q^2_y -q^2_x) -\sin2\phi q_x q_y\right]},
\end{eqnarray} 
where $\theta$ is measured from the normal direction to the plane formed by two axes of $k_1$ and $k_2$ valleys. The contribution of this scattering to spin relaxation is negligible and can be seen analytically by substituting the square-root expression in Eq.~(\ref{eq:U_tail_long}) by $q$ (since both are of the same of order of magnitude). The resulting spin relaxation of a conduction electron with energy $\varepsilon$ above the band edge is
\begin{equation}
\frac{1}{\tau_{s,i}(k)} = \frac{2N_d}{\pi}\frac{\hbar^2}{m_e a_{B}^{2}}\left(\frac{P\Upsilon_{so} }{ E^2_g}\right)^2 \frac{1}{\sqrt{2m_e\varepsilon}} \left[ \ln{\frac{\xi_{\lambda} - 1}{\xi_{\lambda} + 1}} + \frac{2}{\xi_{\lambda}-1} \right], \qquad \xi_{\lambda} = 1 + \frac{\hbar^2}{4m_e\varepsilon\lambda^2} , \qquad \lambda = \sqrt{\frac{k_BT}{4\pi e^2 N_d}}.
\end{equation}
Assigning $\varepsilon = k_BT$, we get that the resulting spin lifetime is of the order of 1~ms at room temperature for $N_d \sim 10^{19}$~cm$^{-3}$.  Its effect is many orders of magnitude weaker than that by the central cell part, and therefore was ignored in Fig.~2(c).

\vspace{2mm}

\end{widetext}

\end{document}